\documentstyle[aps,preprint,epsfig]{revtex}

\begin {document}
\title
{A DOUBLE PARTON SCATTERING BACKGROUND TO HIGGS BOSON PRODUCTION AT THE LHC}

\author{A. Del Fabbro and D.Treleani} 

\address{ Universit\`a di Trieste; Dipartimento di Fisica Teorica, 
Strada Costiera 11, Miramare-Grignano, and
INFN, Sezione di Trieste, I-34014 Trieste, Italy.}

\maketitle

\begin{abstract}
The experimental capability of recognizing the presence of $b$ quarks in 
complex
hadronic final states has addressed the attention towards 
final states with $b\bar{b}$ pairs for observing
the production of the Higgs boson at the LHC, in the intermediate Higgs mass 
range.
We point
out that double parton scattering processes are going to represent a sizeable
background to the process.
\end{abstract}
\vspace{3cm}

E-mail delfabbr@ts.infn.it \\

E-mail daniel@trieste.infn.it \\

\newpage

\section{Introduction}
The problem of identifying the most convenient signatures for detecting the
Higgs boson production at the LHC has been widely discussed in the 
literature\cite{higgs}.
Most results are summarized in ref.\cite{Moretti}, where, in addition, various 
different
backgrounds to the process are estimated. The $b{\bar b}$ channel is
the most favorite Higgs decay mode when the Higgs mass is below the $W^+W^-$
threshold\cite{higgs}. The confidence in the capability of identifying
efficiently the
$b$ quark jets\cite{bquarks} has therefore addressed towards the detection of 
$b{\bar b}$ pairs
to observe the Higgs boson production at the LHC, if the Higgs mass is in the
range $80GeV<M_H<150GeV$. To reduce the huge QCD background to the $b{\bar b}$ 
pair
production, the $b{\bar b}$ pair is detected in association with an isolated
lepton from the decay of a $W$ boson. The process of
interest to detect the Higgs boson production through the $b{\bar b}$ decay
channel is therefore:
$p+p\to WH+X$, with $W\to l\nu_l$, $H\to b\bar{b}$,
where $l=e,\mu$. 

The purpose of the present note is to point out that 
the same $l,b\bar{b}$ final state can be produced also by a different 
mechanism, namely by
a double parton collision process,
which represents therefore a further background to be taken into account in
addition to the other background processes previously considered. In fact as a 
result of
the
present analysis we find that double parton scatterings may represent
a rather sizeable source of background.

The possibility of hadronic interactions with double parton scattering
collisions was foreseen on rather general grounds long
ago\cite{double}. The process has been recently observed by CDF
\cite{cdf}: 
In a hadronic interaction with a double
parton scattering two different pairs of partons interact independently 
at different points in the transverse space, in the same inelastic hadronic
event. The process is induced by unitarity and, as a consequence, it has
been considered mostly in the regime where the partonic cross sections become
comparable to the total inelastic hadronic cross section, namely large c.m.
energy in the hadronic interaction and relatively low transverse momenta of 
the produced partons. Those are in fact also the conditions where the process 
was observed\cite{cdf}.
In such a kinematical regime one does not expect strong initial state 
correlations in the fractional
momenta of the partons undergoing the double collision process and, with this
simplifying hypothesis, the double parton scattering cross section is
proportional to the product of two single scattering cross
sections. All the new non perturbative information on the structure of the
colliding hadrons provided by the process, in the specific case the
information on the two-body parton correlation in transverse space, reduces to
a scale factor with dimensions of a cross section (the 'effective cross
section'\cite{seff}). In the case of two identical parton interactions, as for
producing four large
$p_t$ jets, the double parton scattering cross section assumes therefore the
simplest factorized form
\begin{equation}
\sigma_D(Jets)=\frac{1}{2}\frac{\sigma_J^2}{\sigma_{eff}}
\label{sigmad}
\end{equation}
$\sigma_J$ is the usually considered single parton scattering cross
section:
\begin{equation}
\sigma_J=\sum_{ff'}\int_{p_t>p_t^{min}} dxdx'd^2p_t G_f(x)G_{f'}(x')\frac
{d\hat{\sigma}_{ff'}}{d^2p_t}
\label{sigmas}
\end{equation}
where $G_f(x)$ is the parton distribution as a function of the momentum 
fraction
$x$ and at the scale $p_t$. The different species of interacting 
partons are indicated with the label $f$ and $d\hat{\sigma}_{ff'}/d^2p_t$ is 
the elementary 
partonic cross section. 
$\sigma_{eff}$ is the effective
cross section and it enters as a 
simple proportionality
factor in the integrated inclusive cross section for a
double parton scattering $\sigma_D$.
The value of $\sigma_{eff}$ represents therefore the whole
output of the measure of the double
parton scattering process in this simplest scheme, which on the other hand has 
shown
to be in agreement with the available experimental evidence\cite{cdf}.
In the 
case of two distinguishable parton
scatterings $A$ and $B$ the factor $1/2$ in Eq.\ref{sigmad} is missing and one 
correspondingly writes
\begin{equation}
\sigma_D(AB)=\frac{\sigma_A\sigma_B}{\sigma_{eff}}
\label{sigmaab}
\end{equation}
The effective cross section 
is a geometrical
property of the hadronic interaction, related to the overlap of the matter 
distribution of the two
interacting hadrons in transverse space. The
expectation is that it is
independent on the c.m. energy of the hadronic collision and on the cutoff 
$p_t^{cut}$\cite{seff}.
Moreover, although
one may expect that different kinds of partons may be distributed in different
ways in the transverse space, one does not expect a strong dependence of
$\sigma_{eff}$ on the different possible partonic reactions. The
simplest possibility to consider is therefore the one where the scale factors 
in
Eq.\ref{sigmad} and in Eq.\ref{sigmaab} are the same. 

In the intermediate Higgs mass range 
the partonic center of mass energy needed for producing the Higgs boson is 
relatively low, as compared to
the overall energy involved in the hadronic collision at the LHC, and one may 
therefore expect that the
factorization in Eq.\ref{sigmaab} may still be a good approximation when
producing partonic states with values of the invariant mass of the order of
the mass of the Higgs. We will
therefore estimate the double parton scattering background to the process
$p+p\to WH+X$, with $W\to l\nu_l$ and $H\to b\bar{b}$,
by using the simplest expression in
Eq.\ref{sigmaab}. We will also take the attitude of considering 
the value of $\sigma_{eff}$ as a universal property of all
double parton interactions and we will use the actual value which was measured
by CDF\cite{cdf}. In this respect
one has to point out that
in the
experimental analysis of CDF the measure of the double parton scattering cross
section has been performed by removing all triple parton collision
events
from the sample of inelastic events with double parton scatterings. The double
parton scattering
cross section measured in the experimental analysis does not
correspond therefore to the inclusive cross section written here above and 
usually
considered in the literature, which allows the simple
inverse proportionality relation between $\sigma_D$ and $\sigma_{eff}$. The
double parton scattering cross section measured by the CDF experiment is
in fact smaller as compared to the double parton scattering cross section
discussed here. As a consequence
the resulting value of the effective cross section,
$\sigma_{eff}|_{CDF}$, is somewhat larger\cite{seff} with respect to 
the quantity suitable to the actual purposes.
By using in the present note $\sigma_{eff}|_{CDF}$ as a scale
factor for the double parton scattering process, we will therefore 
underestimate
the size of 
the background due to double parton
scatterings to the Higgs boson production.
 
\section{Double Parton Scattering Background Process}

A background to the process 
$p+p\to WH+X$, with $W\to l\nu_l$, $H\to b\bar{b}$ is represented
by the double parton scattering interaction where the
intermediate vector boson $W$ and the $b\bar{b}$ pair are 
produced in two independent parton interactions. The corresponding integrated
rate is easily evaluated by combining the expected cross sections 
for $W$ and $b\bar{b}$
production at LHC energy with $\sigma_{eff}$ as in Eq.\ref{sigmaab}.
If one uses $\sigma(W)\times BR(W\to l\nu_l)\simeq40nb\nonumber$ \cite{mrs99}, 
$\sigma(b\bar{b})\simeq 5\times10^2\mu$ and as a value for the scale factor 
\begin{equation}
\sigma_{eff}\,=\,14.5mb
\label{wzh}
\end{equation}
(the observed value is 
$\sigma_{eff}|_{CDF}=14.5\pm1.7^{+1.7}_{-2.3}mb$\cite{cdf}) one obtains
that the cross section for
a double parton collision producing a $W\to l\nu_l$ and a $b\bar{b}$ pair, is 
of
the order of $1.4nb$. The Higgs production cross sections, 
$p+p\to WH+X$, with $W\to
l\nu_l$, $H\to b\bar{b}$, has been estimated to be rather of order of 
$1pb$\cite{Moretti}. By integrating the double parton
scattering cross section over the whole possible configurations of the 
$b\bar{b}$
pair one then obtains a cross section three
orders of magnitude larger that the expected signal from Higgs decay. 
Obviously, rather than the
integrated cross sections, one is interested in
comparing the two differential cross sections as a function of the invariant 
mass of the $b{\bar b}$ pair.

In the calculations of the background and signal  we used, for the matrix elements, the
packages MadGraph \cite{mad} and HELAS\cite{helas}, and the integration was
performed by VEGAS\cite{vegas} with the parton distributions MRS99\cite{mrs99}. 
The cross section to produce
$WH$, followed by $W\to l\nu_l$, $H\to b\bar{b}$, is 
plotted in fig.1 for three different possible values of the Higgs mass, 
and it is compared with 
the double parton scattering cross section $d\sigma_D/d M_{b\bar{b}}$ 
as a function of the invariant mass of the
$b\bar{b}$ system. The estimated signal of Higgs boson production in the 
invariant mass of the $b{\bar b}$ pair\cite{Moretti} 
corresponds to the three possible values for the mass of the Higgs boson, $80$,
$100$ and $120GeV$. The curves refer to background double parton
scattering process. The dashed line is obtained by estimating the cross
section for $b{\bar b}$ production at the lowest order in $\alpha_S$ and by
using as a scale factor in $\alpha_S$ the transverse mass of the $b$ quark. 
The continuous line is a rescaling of the lowest order result by a factor $1.8$
and it corresponds to 
the expectations of the order $\alpha_S^3$ estimates of the $b{\bar b}$
cross section according with ref.\cite{mnr}. The estimated background from 
double parton
scatterings is therefore a factor four of five larger than the expected signal.

In fig.2 we compare the signal and the background after applying all the
typical cuts considered to select the Higgs signal: 

- for the lepton: 
$p_t^l>20$ GeV, $|\eta^l|<2.5$ and isolation from the $b$'s,
$\Delta R_{l,b}>.7$

- for the two $b$ partons: 
$p_t^b>15$ GeV, $|\eta^b|<2$ and $\Delta
R_{b,\bar b}>.7$

As in the previous figure the Higgs signal in the $b{\bar b}$ invariant mass
corresponds to three possible values for the mass of the Higgs boson, $80$,
$100$ and $120GeV$. The dotted line is the single parton scattering
background, where the $Wb{\bar b}$ state is produced directly in a single
partonic interaction. The dashed line is the expected background originated by 
double parton
scattering process evaluated by estimating the $b{\bar b}$ production cross
section at ${\cal O}(\alpha_S^3)$. 
The continuous line is the total expected background.

Fig.2 summarizes our result: also after using the more realistic cuts just
described, the double parton scatterings process remains a rather
substantial component of the background. The difference with the conventional
estimate of the background is immediately evident comparing the total 
background
estimate (continuous curve) with single scattering background (dotted curve).

\section{Conclusions}

In the present note we have discussed the background induced by double 
parton collisions to the detection of the Higgs in the $Wb{\bar b}$ channel.
The large rate of 
production of $b\bar{b}$ pairs expected at the LHC
(the corresponding cross section is of order of $500\mu b$) gives rise to a
relatively large probability of production of a $b\bar{b}$ pair in the
process underlying the $W$ production.
As a
consequence a
very promising channel to detect the production of the Higgs boson, in the
intermediate Higgs mass range, namely the final state with a $b\bar{b}$ pair 
and
with an isolated lepton, is affected by a sizable background due to
double parton collision processes. Although the double parton collision
cross section is a decreasing function of the invariant masses of the $b{\bar 
b}$ pair, the relatively large value of the invariant
mass required to the $b\bar{b}$ pair to be assigned to the Higgs decay is not
large enough, at LHC energies, to allow one to neglect the double parton 
scattering background.

It is rather obvious that the considerations above are not limited to the
$Wb{\bar b}$ channel. Similar arguments can be repeated in several
other cases. In addition to the obvious case of the $Zb{\bar b}$ channel, a 
few examples where we expect that multiple parton scattering processes might 
give a non
secondary effect are the following:
\begin{itemize}
\item{ } $W+{\rm jets}$, $Wb+{\rm jets}$ and $Wb{\bar b}+{\rm jets}$, 
\item{ } $t{\bar t}\to llb{\bar b}$, 
\item{ } $t{\bar b}\to b{\bar b}l\nu$,
\item{ } $b{\bar b}+{\rm jets}$,
\item{ } production of many jets when $p_t^{min}\simeq25 GeV$
\end{itemize}

Although after the cuts the ratio signal to background is still a favorable
number, the actual case discussed here shows that an evaluation of the
background, without keeping into account the contribution of the double parton
collision preocesses, may be rather unrealistic at the LHC and that, in some
cases, the cuts to the final state
considered so far are likely to be rediscussed.

\vskip.25in
{\bf Acknowledgments}

\vskip.25in

We thank Stefano Moretti for many useful discussions and Giuseppe Ridolfi for
the code to evaluate the $b{\bar b}$ cross section at ${\cal O}(\alpha_S^3)$. 
This work was partially supported by the Italian Ministry of University and of
Scientific and Technological Research by means of the Fondi per la Ricerca
scientifica - Universit\`a di Trieste.

\begin{figure}
\vspace{2 cm}
\centerline{
\epsfysize=15cm \epsfbox{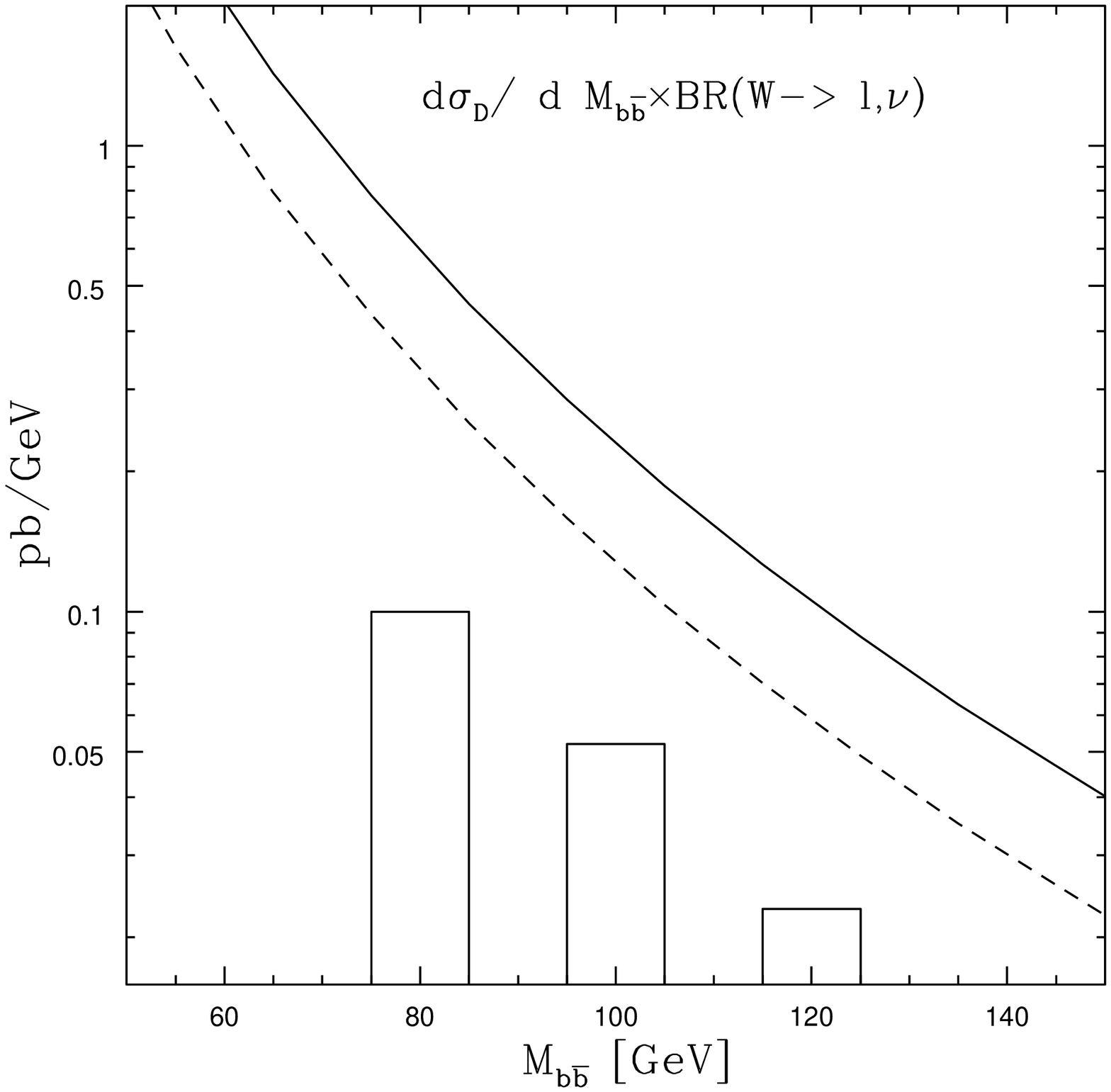}
}
\vspace{1 cm}
\caption[ ]{Double parton scattering background to the Higgs boson production
as a function of the $b{\bar b}$ invariant mass compared to the expected Higgs
signal for three possible values of the Higgs mass, 80, 100 and 120 $GeV$. 
The dashed line is the background at the lowest order in perturbation theory.
The continuous line is the result for the double parton scattering background
when computing the $b{\bar b}$ cross section at order $\alpha_S^3$\cite{mnr}}
\label{tth}
\end{figure} 
\begin{figure}
\vspace{2 cm}
\centerline{
\epsfysize=15cm \epsfbox{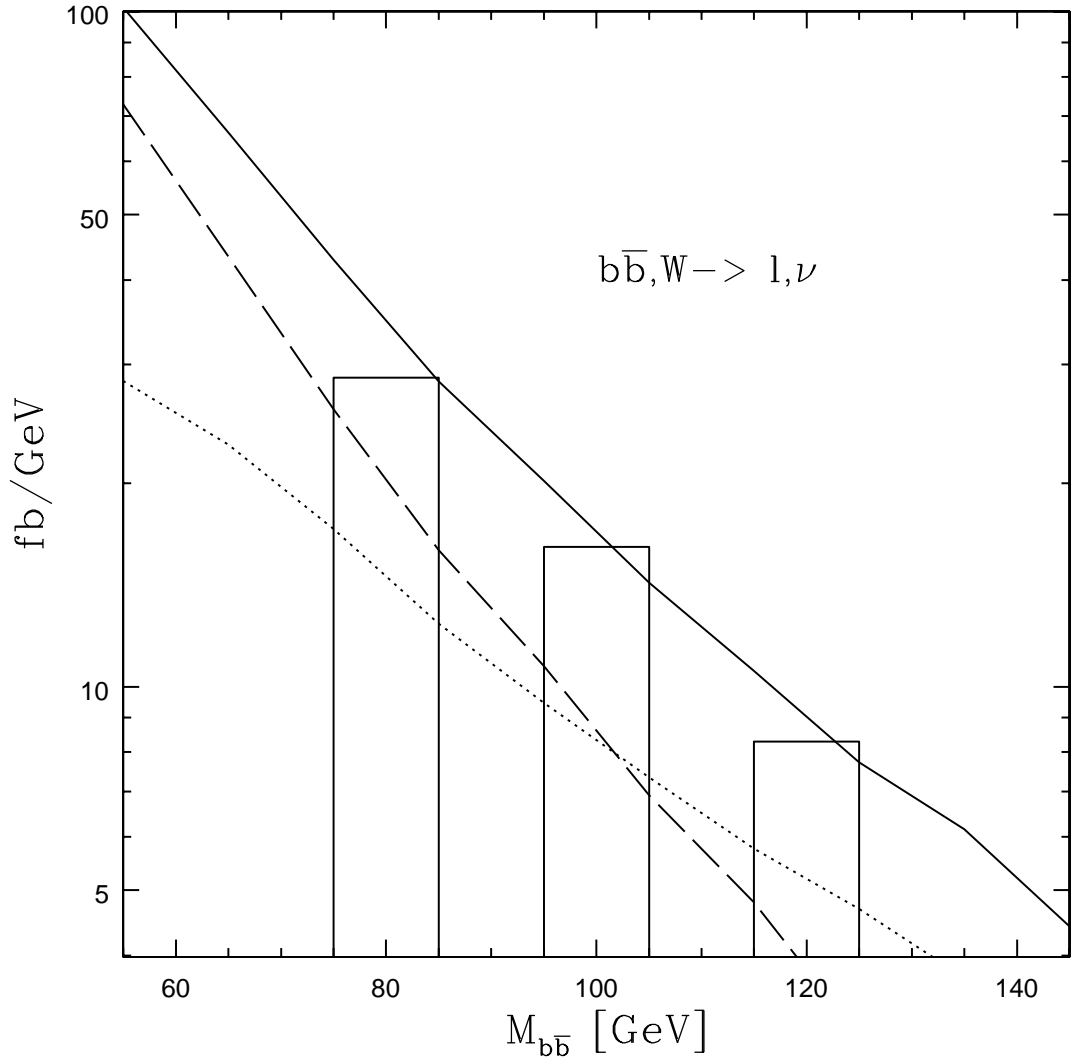}
}
\vspace{1 cm}
\caption[ ]{Backgrounds to Higgs production after the cuts (see main text). 
Dotted line: 
single scattering contribution to the $Wb{\bar b}$ channel. Dashed line:
double parton scattering background. 
Continuous line: total estimated background.} 
\label{tth}
\end{figure}

\end{document}